\documentclass[runningheads]{llncs}
\usepackage[T1]{fontenc}
\usepackage{graphicx}
\usepackage{amsmath}
\usepackage{subcaption}
\newtheorem{assumption}{Assumption}
\usepackage{amsfonts}
\usepackage{amssymb}
\usepackage{xcolor}
\usepackage{color}
\usepackage{amsmath}
\usepackage{amsfonts}
\usepackage{amssymb}
\usepackage{esint}
\newcommand{\ie}{\textit{i.e.}, }

\begin{document}
\begin{sloppypar}

\title{Controlling network-coupled neural dynamics with nonlinear network control theory
\thanks{${\dagger}$ Co-first authors: Zhongye Xia, Weibin Li}
\thanks{Corresponding author: Quanying Liu (liuqy@sustech.edu.cn)}
\thanks{ This work was funded in part by the National Key R\&D Program of China (2021YFF1200804), Shenzhen Science and Technology Innovation Committee (2022410129, KCXFZ2020122117340001).}}
%
\author{Zhongye Xia\inst{1}$\dagger$ \and
Weibin Li\inst{1}$\dagger$ \and
Zhichao Liang\inst{1} \and
Kexin Lou\inst{1} \and Quanying Liu\inst{1}}
%
\authorrunning{Z. Xia et al. (2024)}
%
\institute{Southern University of Science and Technology, Shenzhen, China
}

\maketitle              
\begin{abstract}

This paper addresses the problem of controlling the temporal dynamics of complex nonlinear network-coupled dynamical systems, specifically in terms of neurodynamics. Based on the Lyapunov direct method, we derive a control strategy with theoretical guarantees of controllability. To verify the performance of the derived control strategy, we perform numerical experiments on two nonlinear network-coupled dynamical systems that emulate phase synchronization and neural population dynamics. The results demonstrate the feasibility and effectiveness of our control strategy.

\keywords{Neural Dynamics \and Network Control \and Controllability.}
\end{abstract}
\section{Introduction}



Many natural and man-made systems can be characterized as assemblies of complex network-coupled dynamical systems~\cite{sussillo2014neural,ye2023explainable}. 
Manipulating temporal dynamics of these nonlinear network-coupled dynamical systems has attracted widespread attention, such as synchronizing neural dynamics within resting-state networks~\cite{ponce2015resting,zheng2022kuramoto} and desynchronizing seizure dynamics~\cite{hemami2019numerical,Liang2023Online}.

In neuroscience, electrical stimulation to the brain is an emerging technique to control neural dynamics to achieve synchronization or de-synchronization, which has great potential for treating neural disorders~\cite{yang2021modelling,lou2024data,wang2023multi}. Previous studies have used empirical methods to select stimulation parameters, such as frequency and intensity~\cite{cao2023state}, and linear optimal control theory-based methods to obtain optimal control strategies~\cite{yang2021modelling,liang2024reverse}. However, these approaches hardly achieve the desired outcomes due to the inherent nonlinearity and complexity of the brain network dynamics. It calls for more advanced control strategies with theoretical guarantees to guide the neurostimulation~\cite{Chen2022Pinning,Vega2019InverseOptimal}.  

In this study, we present a Lyapunov-based approach to minimize the error dynamics of each node in a complex nonlinear network. To apply this approach, the system must meet only two simple conditions, \ie Lipschitz continuity and quadratic condition~\cite{zemouche2013lmi,chen2007robust}, ensuring the broad applicability of our controller.
The main contributions are summarized as follows.
\begin{itemize}
  \item By employing the Lyapunov direct method, the proposed control strategy provides a theoretical guarantee of stability and effectiveness.
  \item By conducting numerical experiments on complex nonlinear network-coupled dynamics, such as the Jansen-Rit model and Kuramoto oscillator, we demonstrate the efficacy and robustness of our control strategy in diverse scenarios.
\end{itemize}

                                           
\section{Problem statement}

Given a nonlinear dynamical system with $n$ nodes and $p$ state variables for each node, the inherent dynamics of each node $\mathbf{x}_{i}\in \mathbb{R}^{p}$ and the reference dynamics $\mathbf{x}_{r}\in \mathbb{R}^{p}$ are described as follows,
\begin{align}\label{eq:network_dynamics}
    &\text{Inherent dynamics: } \:\:\Dot{\mathbf{x}}_{i} = f(\mathbf{x}_{i}) - c\displaystyle\sum_{j =1}^{n} L_{ij}\operatorname{h}(\mathbf{x}_{j})+{u}_{i}, \\
    &\text{Reference dynamics: } \Dot{\mathbf{x}}_{r} =f({\mathbf{x}}_{r}) - c\displaystyle\sum_{j =1}^{n}{L}_{ij}^{}\operatorname{h}(\mathbf{x}_{r}),
\end{align}
where $f(\cdot), h(\cdot)$ : ${\mathbb{R}}^{p}\xrightarrow[]{}{\mathbb{R}}^{p}$, $f(\cdot)$ determines the system dynamical properties; $h(\cdot)$ depicts the coupling relationship between nodes; $c$ is the coupling strength between the interconnected nodes; $L_{ij}$ are elements of the Laplacian matrix ${L}$. 

The control goal is to track the reference dynamics by the control input ${u}$. The control error of node $i$ is measured by the difference between inherent and reference dynamics,
${e}_{i} = \mathbf{x}_{i} -  \mathbf{x}_{r}$.
The control input ${u}_{i}$ is designed as
${u}_{i}=-{w}_{i}\varPsi {e}_{i}$,
where ${w}_{i}$ = 0 or 1 determines whether the node is controlled and $\varPsi$ is the control gain.

\section{Analytical results}
To verify the controllability of the proposed strategy, we first outline the assumptions regarding the network-coupled dynamical system~\cite{della2023nonlinear}. 
\begin{assumption}
\label{assumption1}
    $f(\cdot)$ satisfy the quadratic condition around $\mathbf{x}_{r}$ which requires the existence of a scalar ${\theta}_{f}$ such that, $\forall z\in {\mathbb{R}}^{p}$ :
\begin{equation}
    {z}^{T}\left \lgroup f({z}+\mathbf{x}_{r})-f(\mathbf{x}_{r}) \right \rgroup\leqslant{\theta}_{f}{z}^T{z}
\end{equation}
\end{assumption}

\begin{assumption}
\label{assumption2}
    $h(\cdot)$ is globally Lipschitz with constant ${\theta}_{h}\geqslant 0$ means $\forall y, z \in{\mathbb{R}}^{p}$ 
\begin{equation}
    \left \| h(z)-h(y)\right \|\leqslant{\theta}_{h}\left \| {z}-{y}\right \|
\end{equation}
\end{assumption}


\begin{theorem}
\label{theorem1}
    Let Assumptions~\ref{assumption1} \&~\ref{assumption2} hold, if ${\lambda}_{max}\leqslant0$, where ${\lambda}_{max}$ is the largest eigenvalue of $\left \lgroup\left \lgroup{\theta}_{f}+c{\theta}_{h} \left \| {L} \otimes {I}_{p}\right \|\ \right \rgroup{I}_{n}-\Phi{W}_{n}\right \rgroup\otimes{I}_{p}$.
The network dynamics of Eq.~\ref{eq:network_dynamics} is controlled onto $\mathbf{x}_r(t)$.
\end{theorem}

\begin{proof}
Considering the following Lyapunov function $\operatorname{V}(e) = \displaystyle\sum_{i=1}^{n}\frac{1}{2}{e}_{i}^{T}{e}_{i}.$

The Lyapunov function $\operatorname{V}(e)$ describes the stability of the system, and then we get the derivative of the Lyapunov function:
\begin{equation}
\Dot{\operatorname{V}}(e)=\displaystyle\sum_{i=1}^{n}{e}_{i}^{T}\Dot{{e}}_{i},
\end{equation}
where the derivative of the error for each node $\Dot{{e}}_{i}$:
\begin{equation}
\Dot{{e}}_{i}= f({x}_{i})-f({x}_{r})+c\displaystyle\sum_{j =1}^{n}{L}_{ij}\left \lgroup \operatorname{h}({x}_{r})-\operatorname{h}({x}_{j})\right \rgroup +{u}_{i}.
\end{equation}
We can obtain $\Dot{\operatorname{V}}(e)$:
\begin{equation}
    \Dot{\operatorname{V}}(e)=\displaystyle\sum_{i=1}^{n}{e}_{i}^{T}\left \lgroup f({x}_{i})-f({x}_{r})+c\displaystyle\sum_{j =1}^{n}{L}_{ij}\left \lgroup \operatorname{h}({x}_{r})-\operatorname{h}({x}_{j}) \right \rgroup+{u}_{i}\right \rgroup.
\end{equation}
$\Dot{\operatorname{V}}(e)$ can be written as $\Dot{\operatorname{V}}(e)={v}_{1}+{v}_{2}+{v}_{3}$,
where 
\begin{equation}
\begin{aligned}
     {v}_{1} &= \displaystyle\sum_{i=1}^{n}{e}_{i}^{T}\left \lgroup f({x}_{i})-f({x}_{r})\right \rgroup, \\
    {v}_{2} &= \displaystyle\sum_{i=1}^{n}{e}_{i}^{T}c\displaystyle\sum_{j =1}^{n}{L}_{ij}\left \lgroup \operatorname{h}({x}_{r})-\operatorname{h}({x}_{j}) \right \rgroup, \\
    {v}_{3} &= \displaystyle\sum_{i=1}^{n}{e}_{i}^{T} {u}_{i}
\end{aligned}
\end{equation}
\end{proof}

\subsubsection{bound for ${v}_{1}$}:
From assumption~\ref{assumption1}
\begin{equation}
    \begin{aligned}
        {v}_{1}&=\displaystyle\sum_{i=1}^{n}{e}_{i}^{T}\left \lgroup f({x}_{i})-f({x}_{r})\right \rgroup
           =\displaystyle\sum_{i=1}^{n}{e}_{i}^{T}\left \lgroup f({x}_{r}+{e}_{i})-f({x}_{r})\right \rgroup\\
           &\leqslant \displaystyle\sum _{i=1}^{n}\left |{e}_{i}^{T}\left \lgroup f({x}_{r}+{e}_{i})-f({x}_{r})\right \rgroup\right |
           \leqslant \displaystyle\sum_{i=1}^{n}\left \| {e}_{i}^{T}\right \|\left \|f({x}_{r}+{e}_{i})-f({x}_{r}) \right \|\\
           &\leqslant \displaystyle\sum_{i=1}^{n}{\theta }_{f}{e}_{i}^{T}{e}_{i}
           ={e}^{T}\left ({\theta }_{f}{I}_{n} \right )\otimes {I}_{p}{e},
    \end{aligned}
\end{equation}
where ${\theta }_{f}$ is a scalar,${I}_{n}$  ${I}_{p}$ are identity matrices of dimensions \(n \times n\) and \(p \times p\), respectively.

\subsubsection{bound for ${v}_{2}$}:
From assumption~\ref{assumption2}
\begin{equation}
    \begin{aligned}
        {v}_{2} &= \displaystyle\sum_{i=1}^{n}{e}_{i}^{T}c\displaystyle\sum_{i =1}^{n}{L}_{ij}\left \lgroup \operatorname{h}({x}_{r})-\operatorname{h}({x}_{j}) \right \rgroup
            = c{e}^{T}\left \lgroup \ {L} \otimes {I}_{p}\right \rgroup{H}\\
            &\leqslant c \left |{e}^{T}\left \lgroup \ {L} \otimes {I}_{p}\right \rgroup{H} \right |
            \leqslant c \left \|{e}^{T} \right \|\left \| {L} \otimes {I}_{p}\right \|\left \|{H} \right \| 
    \end{aligned}
\end{equation}
where $H$ = $\left [ h(x_{r}) - h(x_{1}) ,  \cdots   , h(x_{r}) - h(x_{n})\right ]$.

From assumption~\ref{assumption2}
\begin{equation}
    \begin{aligned}
        \left \|{H} \right \|^2&=\displaystyle\sum_{i=1}^{n}\left \| \operatorname{h}({x}_{r})-\operatorname{h}({x}_{i})\right \|^2
                           \leqslant 
                           {\theta}_{h}^2\displaystyle\sum_{i=1}^{n} \left \|{e}_{i} \right \|^2
                           ={\theta}_{h}^2\left \|{e} \right \|^2 
    \end{aligned}
\end{equation}

Thus, $\left \| {H}\right \| \leqslant {\theta}_{h}\left \|{e} \right \|$. We can derive that:
\begin{equation}
    {v}_{2} \leqslant c{\theta}_{h} \left \|{e}^{T} \right \|\left \| {L} \otimes {I}_{p}\right \|\left \|{e} \right \|
            = {e}^T\left \lgroup c{\theta}_{h} \left \| {L} \otimes {I}_{p}\right \|{I}_{n}\right \rgroup\otimes{I}_{p}{e}
\end{equation}

\subsubsection{bound for ${v}_{3}$}:
\begin{equation}
    {v}_{3}=\displaystyle\sum_{i=1}^{n}{e}_{i}^{T} {u}_{i}
           =-\Phi\displaystyle\sum_{i=1}^{n}{e}_{i}^{T} {w}_{i} {e}_{i}
           =-\Phi{e}^T\left \lgroup {W}_{n}\otimes{I}_{p}\right \rgroup{e},
\end{equation}
where $W_n = \mathrm{diag}(w_1, w_2, \ldots, w_n)$, $diag(\cdot)$ is a notation of creating a diagonal matrix with input elements

\subsubsection{bound for $\Dot{\operatorname{V}}(e)$}:
\begin{equation}
    \begin{aligned}
       \Dot{\operatorname{V}}(e)
                             \leqslant{e}^{T}\left \lgroup\left \lgroup{\theta}_{f}+c{\theta}_{h} \left \| {L} \otimes {I}_{p}\right \|\ \right \rgroup{I}_{n}-\Phi{W}_{n}\right \rgroup\otimes{I}_{p}{e}\leqslant {\lambda}_{max}{e}^{T}{e},
    \end{aligned}
\end{equation}
where ${\lambda}_{max}$ is the largest eigenvalue of Eq.~\eqref{eq:lyapunov_stability}
\begin{equation}
\label{eq:lyapunov_stability}
    \left \lgroup\left \lgroup{\theta}_{f}+c{\theta}_{h} \left \| {L} \otimes {I}_{p}\right \|\ \right \rgroup{I}_{n}-\Phi{W}_{n}\right \rgroup\otimes{I}_{p}.
\end{equation}

If ${\lambda}_{max}\leqslant0$, the controlled network is globally stable about $x_r(t)$.

\section{Numerical experiments on application scenarios}   
Two numerical simulations are given to show the effectiveness of the preceding control strategy in this section. Firstly, consider a double cortical columns Jansen-Rit model that simulates the seizure propagation process from an epileptogenic node to a propagation node. The network-coupled Jansen-Rit model and the parameters of which are summarized in our previous study~\cite{Liang2023Online}.

\begin{figure}[ht]
    \centering
    \includegraphics[width=0.9\linewidth]{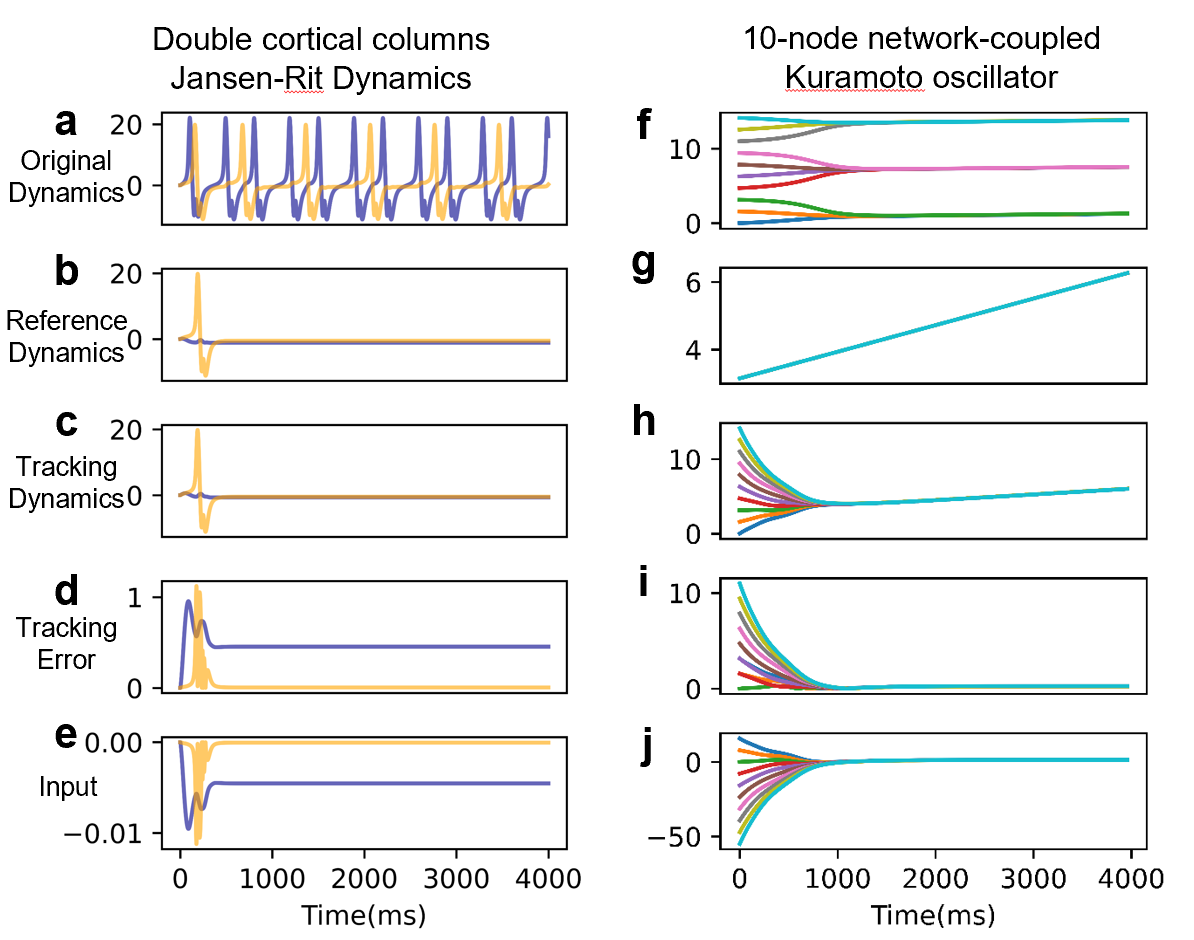}
    \caption{Two examples: (Left) Seizure suppression on the double cortical columns Jansen-Rit model. (Right) Phase synchronization on the network-coupled Kuramoto oscillator.}
    \label{fig:image1}
\end{figure}


The controlled system exhibits a seizure-like behavior (shown in Fig.~\ref{fig:image1}a). The reference dynamics are shown in Fig.~\ref{fig:image1}b. 
By simple calculation, we can verify that Assumption~\ref{assumption1} \&~\ref{assumption2} hold with $\theta_f \approx 27$ and $\theta_h \approx 0.0025$. Based on Theorem~\ref{theorem1}, the controlled system followed the reference system if the largest eigenvalue of Eq.~\eqref{eq:lyapunov_stability} satisfied ${\lambda}_{max}\leqslant0$. Thus, we take $\Phi = 30$ in our controller for brief. Fig.~\ref{fig:image1}c shows the temporal dynamics under the proposed control strategy and Fig.~\ref{fig:image1}d shows the evolution of tracking error relative to the reference dynamics. The result shows that the controlled system converged to the reference dynamics with inputs (Fig.~\ref{fig:image1}e).

Further, we conduct the numerical simulations on the network-coupled Kuramoto oscillator with $N=10$ nodes:
\begin{equation}
\label{eq:kuramoto_dynamics}
\dot{\theta}_i = \omega_i + \frac{K}{N} \sum_{j=1}^{N} \sin(\theta_j - \theta_i), \quad i = 1, \ldots, N
\end{equation}

The temporal dynamics of the controlled system~\eqref{eq:kuramoto_dynamics} is shown in Fig~\ref{fig:image1}f with parameters $\omega_i=(i-1)\pi/2, K=10$. The reference dynamics are generated with parameters $\omega_i= \pi/2, K=0$ (Fig.~\ref{fig:image1}g).
With numerical analysis, Assumption~\ref{assumption1} \&~\ref{assumption2} hold with $\theta_f = 0$ and $\theta_h = 1$. Based on Theorem~\ref{theorem1}, the controlled system followed the reference dynamics if the largest eigenvalue of Eq.~\eqref{eq:lyapunov_stability} satisfied ${\lambda}_{max}\leqslant0$. We choose $\phi = 1.5$ in our controller. Fig.~\ref{fig:image1}h shows that the dynamics of the controlled system followed the reference dynamics with small deviation controlled with calculated input (see Fig.~\ref{fig:image1}i,j).


\section{Conclusion}
In this study, we presented a robust control strategy for manipulating the temporal dynamics of nonlinear network-coupled dynamical systems, and we demonstrated its efficacy through theoretical analysis and numerical experiments. Our approach facilitates the development of tailored control policies for personalized neurostimulation. 


\bibliographystyle{IEEEtran}
\bibliography{mybib}

\end{sloppypar}
\end{document}